\documentstyle[12pt]{article}

\begin{document}

\newcommand{\ra}{\rightarrow}
\newcommand{\ko}{K^0}
\newcommand{\be}{\begin{equation}}
\newcommand{\ee}{\end{equation}}
\newcommand{\bea}{\begin{eqnarray}}
\newcommand{\eea}{\end{eqnarray}}
\def\dop{{\rm d}\hskip -1pt}
\def\bfone{\relax{\rm 1\kern-.35em 1}}
\def\bfzero{\relax{\rm I\kern-.18em 0}}
\def\inbar{\vrule height1.5ex width.4pt depth0pt}
\def\IC{\relax\,\hbox{$\inbar\kern-.3em{\rm C}$}}
\def\ID{\relax{\rm I\kern-.18em D}}
\def\IF{\relax{\rm I\kern-.18em F}}
\def\IK{\relax{\rm I\kern-.18em K}}
\def\IH{\relax{\rm I\kern-.18em H}}
\def\II{\relax{\rm I\kern-.17em I}}
\def\IN{\relax{\rm I\kern-.18em N}}
\def\IP{\relax{\rm I\kern-.18em P}}
\def\IQ{\relax\,\hbox{$\inbar\kern-.3em{\rm Q}$}}
\def\IR{\relax{\rm I\kern-.18em R}}
\def\IG{\relax\,\hbox{$\inbar\kern-.3em{\rm G}$}}
\font\cmss=cmss10 \font\cmsss=cmss10 at 7pt
\def\ZZ{\relax\ifmmode\mathchoice
{\hbox{\cmss Z\kern-.4em Z}}{\hbox{\cmss Z\kern-.4em Z}}
{\lower.9pt\hbox{\cmsss Z\kern-.4em Z}}
{\lower1.2pt\hbox{\cmsss Z\kern-.4em Z}}\else{\cmss Z\kern-.4em
Z}\fi}
\def\a{\alpha} \def\b{\beta} \def\d{\delta}
\def\e{\epsilon} \def\c{\gamma}
\def\G{\Gamma} \def\l{\lambda}
\def\L{\Lambda} \def\s{\sigma}
\def\cA{{\cal A}} \def\cB{{\cal B}}
\def\cC{{\cal C}} \def\cD{{\cal D}}
\def\cF{{\cal F}} \def\cG{{\cal G}}
\def\cH{{\cal H}} \def\cI{{\cal I}}
\def\cJ{{\cal J}} \def\cK{{\cal K}}
\def\cL{{\cal L}} \def\cM{{\cal M}}
\def\cN{{\cal N}} \def\cO{{\cal O}}
\def\cP{{\cal P}} \def\cQ{{\cal Q}}
\def\cR{{\cal R}} \def\cV{{\cal V}}\def\cW{{\cal W}}
%
%
%
\def\crr{\crcr\noalign{\vskip {8.3333pt}}}
\def\tilde{\widetilde}
\def\bar{\overline}
\def\us#1{\underline{#1}}
\let\shat=\hat
\def\hat{\widehat}
\def\hyp{\vrule height 2.3pt width 2.5pt depth -1.5pt}
\def\square{\mbox{.08}{.08}}
\def\Coeff#1#2{{#1\over #2}}
\def\Coe#1.#2.{{#1\over #2}}
\def\coeff#1#2{\relax{\textstyle {#1 \over #2}}\displaystyle}
\def\coe#1.#2.{\relax{\textstyle {#1 \over #2}}\displaystyle}
\def\half{{1 \over 2}}
\def\shalf{\relax{\textstyle {1 \over 2}}\displaystyle}
\def\dag#1{#1\!\!\!/\,\,\,}
\def\to{\rightarrow}
\def\notin{\hbox{{$\in$}\kern-.51em\hbox{/}}}
\def\shdot{\!\cdot\!}
\def\ket#1{\,\big|\,#1\,\big>\,}
\def\bra#1{\,\big<\,#1\,\big|\,}
\def\equaltop#1{\mathrel{\mathop=^{#1}}}
\def\Trbel#1{\mathop{{\rm Tr}}_{#1}}
\def\inserteq#1{\noalign{\vskip-.2truecm\hbox{#1\hfil}
\vskip-.2cm}}
\def\attac#1{\Bigl\vert
{\phantom{X}\atop{{\rm\scriptstyle #1}}\phantom{X}}}
\def\exx#1{e^{{\displaystyle #1}}}
\def\del{\partial}
\def\delbar{\bar\partial}
\def\nex#1{$N\!=\!#1$}
\def\dex#1{$d\!=\!#1$}
\def\cex#1{$c\!=\!#1$}
\def\eg{{\it e.g.}} \def\ie{{\it i.e.}}
%
\def\cS{{\cal K}}
\def\IE{\relax{{\rm I\kern-.18em E}}}
\def\cE{{\cal E}}
\def\rt{{\cR^{(3)}}}
\def\IGam{\relax{{\rm I}\kern-.18em \Gamma}}
\def\IGa{\IA}
\def\LG{Lan\-dau-Ginz\-burg\ }
\def\cV{{\cal V}}
\def\Rt{{\cal R}^{(3)}}
\def\wabc{W_{abc}}
\def\WABC{W_{\a\b\c}}
\def\W{{\cal W}}
\def\tft#1{\langle\langle\,#1\,\rangle\rangle}
\def\IA{\relax{\hbox{{\rm A}\kern-.82em {\rm A}}}}
\let\picfuc=\fp
\def\hata{{\shat\a}}
\def\hatb{{\shat\b}}
\def\hatA{{\shat A}}
\def\hatB{{\shat B}}
\def\bv{{\bf V}}
\def\K{K\"ahler}
\def\w{w}
\def\CP{C\!P}
\def\o#1#2{{{#1}\over{#2}}}
\newtheorem{definizione}{Definition}[section]
\newcommand{\bd}{\begin{definizione}}
\newcommand{\ed}{\end{definizione}}
\newtheorem{teorema}{Theorem}[section]
\newcommand{\bth}{\begin{teorema}}
\newcommand{\eth}{\end{teorema}}
\newtheorem{lemma}{Lemma}[section]
\newcommand{\blem}{\begin{lemma}}
\newcommand{\elem}{\end{lemma}}
\newcommand{\brr}{\begin{array}}
\newcommand{\err}{\end{array}}
\newcommand{\nn}{\nonumber}
\newtheorem{corollario}{Corollary}[section]
\newcommand{\bcorol}{\begin{corollario}}
\newcommand{\ecorol}{\end{corollario}}
\def\twomat#1#2#3#4{\left(\begin{array}{cc}
 {#1}&{#2}\\ {#3}&{#4}\\
\end{array}
\right)}
\def\twovec#1#2{\left(\begin{array}{c}
{#1}\\ {#2}\\
\end{array}
\right)}

\begin{titlepage}
\hskip 9cm
\vbox{\hbox{hep-th/9611076}\hbox{November, 1996}}
\vfill
\begin{center}
{\large { Supergravities in Diverse Dimensions
and Their Central Extension \footnote{Talk given by
R. D'Auria at the ``Workshop on Gauge Theories,
Applied Supersymmetry and Quantum Gravity'',
Imperial College, London, UK, 5 - 10 July 1996}}}\\
\vskip 1.5cm
  {\bf Laura Andrianopoli$^1$ and
Riccardo D'Auria$^2$ } \\
\vskip 0.5cm
{\small
$^1$ Dipartimento di Fisica, Universit\'a di Genova, via Dodecaneso 33,
I-16146 Genova
and Istituto Nazionale di Fisica Nucleare (INFN) - Sezione di Genova,
Italy\\
\vspace{6pt}
$^2$ Dipartimento di Fisica, Politecnico di Torino,
 Corso Duca degli Abruzzi 24, I-10129 Torino
and Istituto Nazionale di Fisica Nucleare (INFN) - Sezione di Torino,
Italy }
\end{center}
\vfill
\begin{center} {\bf Abstract}
\end{center}
{\small
 In this lecture  moduli dependent charges for $p$--extended objects
 are analyzed
for generic $N$-extended supergravities in dimensions $4 \leq D <10$.
Differential relations and sum rules among the charges are derived.
}
\vspace{2mm} \vfill \hrule width 3.cm
{\footnotesize
\noindent
$^*$ Supported in part by DOE grant
DE-FGO3-91ER40662, Task C.
and by EEC Science Program SC1*CT92-0789.}
\end{titlepage}

In recent time attempts to study non perturbative properties of gauge
theories \cite{sw} and string theories \cite{ht,wit} have
made an essential use of low energy
effective lagrangians incorporating the global and local symmetries
of the fundamental theories.
In this analysis BPS states play an important role \cite{sen, ss},
especially in
connection with enhancement of gauge symmetries \cite{cdfv, ht2, w2}
and more generally
for phase transitions which may be signaled by some BPS state
becoming massless at some point of the underlying moduli space.
 \\
The BPS states often appear as solitonic solutions of the
supergravity field equations in backgrounds preserving some of the
supersymmetries depending on the degree of extremality of the
solitonic state.
Recently a lot of information on black holes and black $p$-branes in
diverse dimensions have been obtained using these methods.
\\
For instance, extremal black holes preserving one
 supersymmetry in $D=4$ and $5$ dimensions  have an entropy formula
obtained in a rather moduli independent way
by minimizing the ADM mass in the moduli space \cite{fk1}.
\par
The underlying geometry of the moduli space plays a fundamental role
in finding these solutions since the ADM mass or, more
generally, the mass per
unit of $p$--volume for $p$--extended objects depends on the
asymptotic value of the moduli and some other physical quantities,
such as the classical determination of the Bekenstein-Hawking
entropy formula, are also related to properties of the moduli space
(for a detailed bibliography on these topics see references in \cite{adf}).
\\
In view of these applications it is interesting to see how far the
formalism used to derive the Bekenstein--Hawking entropy formula
from extremization of central charges can be extended to higher $N$
supergravities in diverse dimensions \cite{sase}.\\
In this lecture we give a short account of the group theoretical
formalism which underlies the construction of central and matter
charges. For an application to the black hole entropy formula, see
ref.
\cite{adf}. A more extended version of the present paper is given in
ref.
\cite{adf2}.
\\
First of all we observe that, in view of several non perturbative
dualities between different
kinds of theories, a given theory is truly specified by the dimension of
space--time in which it lives, the number of unbroken supersymmetries
and the massless matter content.
 \\
With the exception of $D=4$, $N=1,2$ and $D=5$, $ N=2$  all supergravity
theories contain scalar fields whose kinetic Lagrangian is described
by $\sigma$--models of the form $G/H$.
Here $G$ is a non compact group acting as an isometry group on the
scalar manifold while $H$, the isotropy subgroup, is of the form:
\begin{equation}
H=H_{Aut} \otimes H_{matter}
\end{equation}
 $H_{Aut}$ being the automorphism group of the supersymmetry algebra
 while $H_{matter}$ is related to the matter multiplets
 (Of course $H_{matter}=\bfone$ in all cases where supersymmetric matter
 doesn't exist, namely $N>4$ in $D=4,5$ and in general in all
 maximally extended supergravities).
The coset manifolds $G/H$ and the automorphism groups for various
 supergravity theories for any $D$ and $N$  can be found in the literature
 (see for instance the reference book
\cite{sase}).
 As it is well known, the group G acts linearly on the $n=p+2$--forms
 field strengths $H^\Lambda _{a_1\cdots a_n}$ corresponding to the
 various ($p+1$)--forms appearing in the gravitational and matter
 multiplets. Here and in the following the index $\Lambda$ runs over
 the dimensions of some representation of the duality group $G$.
The true duality symmetry, acting on integral  quantized electric
and magnetics charges,
 is  the restriction of  the continuous group $G$ to the integers
 \cite{ht}.
 \\
 All the properties of the given supergravity theories for fixed $D$
 and $N$ are completely fixed in terms of the geometry of $G/H$
 namely in terms of the coset representatives $L$ satisfying the
 relation
 \begin{equation}
g L(\phi) = L(\phi ^\prime) h^{-1} (g,\phi)
\end{equation}
where $g\in G$, $h\in H$  and $\phi ^\prime =   \phi ^\prime
 (\phi)$,
 $\phi$ being the coordinates of $G/H$.
 In particular, as explained in the following, the kinetic metric for
 the ($p+2$)--forms $H^\Lambda$ is fixed in terms of $L$ and the
 physical field strengths of the interacting theories are "dressed"
 with scalar fields in terms of the coset representatives.
 This allows us to write down the central charges associated to the
 ($p+1$)--forms in the gravitational multiplet in a neat way in terms
 of the geometrical structure of the moduli space. \\
In an analogous way also the matter ($p+1$)--forms of the matter
multiplets give rise to charges which, as we will see, are closely
related to the central charges.
\par
Our main goal is to write down the explicit form of the dressed
charges and to find relations among them analogous to those worked
out in $D=4$, $N=2$ case  by means of the Special Geometry relations
\cite{cdf,cdfv}.
\\
To any ($p+2$)--form $H^\Lambda$ we may associate a magnetic
charge ($D-p-4$--brane)
and  an
electric ($p$--brane) charge given respectively by:
\begin{equation}
g^\Lambda = \int _{S^{p+2}} H^\Lambda
\qquad \qquad
e_\Lambda = \int _ {S^{D-p-2}}  \cG _\Lambda
\end{equation}
where $\cG_{\Lambda}= -{\rm i} {\partial \cL \over \partial H^\Lambda}$.
These charges however are not the physical charges of the interacting
theory; the latter ones can be computed by looking at the
transformation laws of the fermion fields, where the physical
field--strengths appear dressed with the scalar fields.
Let us first introduce the central charges:
they are associated to the dressed ($p+2$)--forms $H^\Lambda$ appearing
in the supersymmetry transformation law of the gravitino 1-form.
Quite generally we have, for any $D$ and $N$:
\begin{equation}
\delta \psi_A = D\epsilon_A + \sum_{i} c_i L_{\Lambda_i AB} (\phi)
H^{\Lambda_i} _ {a_1\cdots a_{n_i}}\Delta^{a a_1\cdots a_n}
\epsilon^B V_a+ \cdots
\label{tragra}
\end{equation}
where $
\Delta_{a a_1\cdots a_n}=\left( \Gamma _{a a_1 \cdots
a_{n_i}} - {n \over n-1} (D-n-1)\delta^a_{[a_1} \Gamma_{a_2\cdots
a_{n_i}]} \right)
$.   \\
 Here $c_i$ are coefficients fixed by supersymmetry, $V^a$ is the
 space--time vielbein, $A=1,\cdots,N$ is the index acted on by the
 automorphism group, $\Gamma_{a_1\cdots a_n}$ are $\gamma$--matrices
 in the appropriate dimensions, and the sum runs over all the ($p+2$)--forms
 appearing in the gravitational multiplet. Here and in the following
 the dots denote trilinear fermion terms. $L_{\Lambda AB}$ is given
 in terms of the coset representative matrix of $G$.
 Actually it coincides with a subset of the columns of this matrix
 except in $D=4$ ($N>1$) and the for maximally extended
 $D=6,8$ supergravities since in those cases we have the slight
 complication  that the action of $G$ on the $p+2 =D/2$--forms is
 realized through the embedding of $G$ in $Sp(2n,\IR)$ or $O(n,n)$
 groups.
 Excluding for the moment these latter cases,
$L_{\Lambda AB}$  is actually a set of columnes of the (inverse) coset
representative  $L$ of $G$.
Indeed, let us decompose the representative of $G/H$ as follows:
\begin{equation}
L=(L^\Lambda_{\ AB}, L^\Lambda_{\ I}) \qquad\quad L^{-1} =(L^{AB}_{\
\
\Lambda}, L^I_{\ \Lambda})
\label{defl}
\end{equation}
where the couple of indices $AB$ transform as a symmetric tensor
under $H_{Aut}$
and $I$ is an index in the fundamental
representation of $H_{matter}$ which in general is an orthogonal group
(in absence of matter multiplets
$L\equiv ( L^\Lambda_{\ AB})$).
Quite generally we have:
\begin{equation}
 L_{AB\Lambda}L^{AB}_{\ \ \Sigma} + L_{I\Lambda}L^{I}_{\ \Sigma}=
\cN_{\Lambda\Sigma}\label{ndil}
\end{equation}
where
  $\cN$ defines the kinetic matrix of the $(p+2)$--forms $H^\Lambda$
 and the indices of $H_{Aut}$ are raised
 and lowered with the appropriate
metric in the given representation.
  Note that both for matter coupled and maximally extended
supergravities we have:
$L_{\Lambda AB} = \cN_{\Lambda\Sigma}L^{\Sigma}_{\ AB} $.
\par
For maximally extended supergravities
 $\, \cN_{\Lambda\Sigma} = L_{AB\Lambda}L^{AB}_{\ \ \Sigma}$.
When $G$ contains an orthogonal factor $O(m,n)$, what happens
 for matter coupled supergravities in $D=5,7,8,9$, where $G=O(10-D,n)
 \times O(1,1)$
 and in all the matter coupled and the maximally extended $D=6$ theories,
 the coset
representatives of the orthogonal group satisfy the extra relation:
\begin{equation}
 L_{AB\Lambda}L^{AB}_{\ \ \Sigma} -  L_{I\Lambda}L^{I}_{\ \Sigma}=
\eta_{\Lambda\Sigma} \label{etadil}
\end{equation}
 where
$\eta_{\Lambda\Sigma}$ is the $O(m,n)$ invariant metric.
(In particular,  setting the matter to zero, we have in these cases
$\cN_{\Lambda\Sigma}= \eta_{\Lambda\Sigma}$).\\
Coming back to equation (\ref{tragra}) we see that the dressed
graviphoton $n_i$--forms field strengths are:
\begin{equation}
 T^{(i)}_{AB} = L_{\Lambda_i AB} (\phi) H^ {\Lambda_i}
\end{equation}
The magnetic central charges for BPS saturated ($D-p-4$)--branes
can be now defined
(modulo numerical factors to be fixed in each theory) by integration
of the dressed
field strengths as follows:
\begin{equation}
Z^{(i)}_{(m) AB} = \int _{S^{p+2}}L_{\Lambda_i AB}(\phi) H^{\Lambda_i}=
 L_{\Lambda_i AB}(\phi_0) g^{\Lambda_i}
 \label{carma}
\end{equation}
where $\phi_0$ denote the $v.e.v.$ of the scalar fields, namely
$\phi_0 = \phi(\infty)$ in a given background.
The corresponding electric central charges are:
 \begin{equation}
Z^{(i)}_{(e) AB} = \int _{S^{D-p-2}}L_{\Lambda_i AB}(\phi)
^{\ \star}H^{\Lambda_i}= \int  \cN _{\Lambda_i\Sigma_i}
L^{\Lambda_i}_{\  AB}(\phi)
^{\ \star}H^{\Sigma_i}=
 L^{\Lambda_i}_{\ AB}(\phi_0) e_{\Lambda_i}
 \end{equation}
 These formulae make it explicit that $L^\Lambda_{\ AB}$ and
 $L_{\Lambda AB}$ are related by electric--magnetic duality via the
 kinetic matrix.
Note that the same field strengths (the graviphotons) which
 appear in the gravitino
 transformation laws are also present in the dilatino transformation law:
    \begin{equation}
\delta \chi_{ABC} = \cdots + \sum_{i} c^\prime_i L_{\Lambda_i AB} (\phi)
H^{\Lambda_i} _ {a_1\cdots a_{n_i}}\Gamma^{a_1\cdots a_n}
\epsilon_C + \cdots
\label{tradil}
\end{equation}
so that we get no new charges from (\ref{tradil}).
   \par
   However, the presence of matter multiplets gives rise to "matter
   charges" associated to matter vectors.
 In fact, when vector multiplets are present,
 the matter vector field
 strengths are dressed with the columns $L_{\Lambda I}$
 of the coset element (\ref{defl})
 and they
 appear in the transformation laws of the gaugino fields:
  \begin{equation}
\delta \lambda^I_{A} = \Gamma^a P^I_{AB , i}
\partial_a \phi^i \epsilon^B +
 c L_{\Lambda}^{\ I} (\phi)
F^{\Lambda} _ {ab}\Gamma^{ ab} \epsilon_A  + \cdots
\label{tragau}
\end{equation}
  where  $ P^I_{AB , i}$ is the vielbein of the coset manifold
  spanned by the scalar fields of the vector multiplets and $c$ is a
  constant fixed by supersymmetry (in $D=6$, $N=(2,0)$ and $N=(4,0)$, the
  2--form $F^{\Lambda}_{ab} \Gamma^{ab}$ is replaced by the 3--form
  $F^{\Lambda}_{abc} \Gamma^{abc}$).
In the same way as for central charges, one finds the
magnetic matter charges:
\begin{equation}
  Z_{(m) A} ^{\ I} = \int _{S^{p+2}} L_\Lambda^{\ I} F^\Lambda
   =   L_\Lambda^{\ I} (\phi_0) g^\Lambda
\end{equation}
while the electric matter charges are:
\begin{equation}
Z_{(e) I} = \int _{S^{D-p-2}}L_{\Lambda I}(\phi)
^{\ \star}F^{\Lambda}= \int _{S^{D-p-2}} \cN _{\Lambda\Sigma}
L^{\Lambda}_{\  I}(\phi)
^{\ \star}F^\Sigma =
 L^{\Lambda}_{\ I}(\phi_0) e_{\Lambda}
\end{equation}
 \par
 The important fact to note is that the central charges and matter
 charges satisfy relations and sum rules analogous to those derived
 in $D=4$, $N=2$ using Special Geometry techniques \cite{cdf}.
Indeed, for a general coset manifold we may introduce the
left--invariant 1--form $\Omega=L^{-1} d L$ satisfying the
relation (see for instance \cite{cadafr}):
\begin{equation}
d \Omega + \Omega \wedge \Omega =0 ;\quad
\Omega=\omega^i T_i + P^\alpha T_\alpha
\label{mc}
\end{equation}
  $T_i, T_\alpha$ being the generators of $G$ belonging
  respectively to the Lie
  subalgebra $\IH$ and to the coset space algebra $\IK$ in the
  decomposition
$\IG = \IH + \IK  $,
 $\IG$ being the Lie algebra of $G$. Here $\omega^i$ is the $\IH$
 connection and $P^\alpha$  is the vielbein of $G/H$.
Suppose now we have a matter coupled theory. Then, using
the decomposition above, from
(\ref{mc})  we get:
\begin{equation}
dL^\Lambda_{\ AB} = L^\Lambda_{\ CD} \omega^{CD}_{\ \ AB} +
L^\Lambda_{\ I} P^I_{AB}
\label{cosetmc}
\end{equation}
where $P^I_{AB}$ is the vielbein on $G/H$ and
 $\omega^{CD}_{\ \ AB}$ is the $\IH$--connection in the given
 representation.
It follows:
\begin{equation}
D^{(H)} L^\Lambda_{\ AB}= L^\Lambda_{\ I} P^I_{AB}
\label{dllp}
\end{equation}
where the derivative is covariant with  respect to the
$\IH$--connection  $\omega^{CD}_{\ \ AB}$.
Using the definition of the magnetic dressed charges given in
(\ref{carma}) we obtain:
\begin{equation}
D^{(H)} Z_{AB}= Z_{I} P^I_{AB}
\label{dz}
\end{equation}
 This is a prototype of the formulae one can derive in the various
 cases for matter coupled supergravities.
 To illustrate one possible application of this kind of formulae let
 us suppose that in a given background preserving some number of
 supersymmetries $Z_I=0$ as a consequence of $\delta\lambda^I_A=0$.
 Then we find:
 \begin{equation}
D^{(H)} Z_{AB}=0 \to d(Z_{AB} Z^{AB} )=0
\end{equation}
 that is the square of the
 central charge reaches an extremum with respect to the
 $v.e.v.$ of the moduli fields.
 For the maximally extended supergravities there are no matter
 field--strengths and the previous differential relations become
differential relations between central charges only.
Indeed from the Maurer--Cartan equations we get in this case:
  \begin{equation}
D^{(H)} Z_{AB}= Z^{CD} P_{CD AB}
\end{equation}
 This relation implies that the vanishing of a subset of central
 charges forces the vanishing of the covariant derivatives of some
 other subset.
 Typically, this happens in some
 supersymmetry preserving backgrounds where
 the requirement $\delta\chi_{ABC}=0$ corresponds to the vanishing of
 just a subset of central charges.
 Finally, from the coset representatives relations
 (\ref{ndil}) (\ref{etadil}) it
 is immediate to obtain sum rules for the central and matter charges
 which are
the counterpart of those found in $N=2$, $D=4$ case using Special
Geometry.
 Indeed, let us suppose e.g. that the group $G$ is
 $G=O(10-D,n)\times O(1,1)$, as it
 happens in general for all the minimally extended supergravities in
 $7 \leq D \leq 9$,  $D=6$ type $IIA$ and
 $D=5$, $N=2$.
  The coset representative is now a tensor product $L \to e^\sigma L$, where
  $e^\sigma$ parametrizes the $O(1,1)$ factor.
  We have, from   (\ref{etadil}):
 $
L^t \eta L =\eta
$,
 where $\eta$ is the invariant metric of $O(10-D,n)$ and,
 from  (\ref{ndil}):
 \begin{equation}
e^{-2\sigma}(L^{t} L)_{\Lambda\Sigma} =\cN_{\Lambda\Sigma}.
\end{equation}
 From (\ref{mc}) and the definition of the charges one finds:
 \begin{equation}
Z_{AB} Z^{AB} - Z_I Z^I = g^\Lambda \eta _{\Lambda\Sigma} g^\Sigma
e^{-2\sigma}
\end{equation}
   \begin{equation}
Z_{AB} Z^{AB} + Z_I Z^I = g^\Lambda \cN _{\Lambda\Sigma} g^\Sigma
\end{equation}
 In more general cases analogous relations of the same kind can be
 derived.
\par
Let us now see how these considerations modify
in  the case of extended objects which
can be dyonic, i.e. for $p=(D-4)/2$.
Following Gaillard and Zumino \cite{gz}, for $p$ even ($D$ multiple of 4)
 the duality group
$G$ must have a symplectic embedding in $Sp(2n,\IR)$;
for $p$ odd ($D$ odd multiple of 2),
the duality group is always $O(n,m)$ where $n$, $m$
are respectively the number of
self--dual and anti self--dual ($p+2$)--forms.
 \\
 In $D=4$, $N>2$ we may decompose the vector field--strengths
 in self--dual and
 anti self--dual parts:
 $
\cF^{\mp} = {1\over 2}(\cF\mp {\rm i} ^{\ \star} \cF)
$.
  According to the Gaillard--Zumino construction, $G$ acts on the
  vector $(\cF^{- \Lambda},\cG^{-}_\Lambda)$
  (or its complex conjugate) as a subgroup of
  $Sp(2 n_v,\IR)$ ($n_v$ is the number of vector fields)
with duality transformations interchanging electric and magnetic
 field--strengths:
 \begin{equation}
{\cal S}
\left(\matrix {\cF^{-\Lambda} \cr
\cG^-_\Lambda\cr}\right)=
\left(\matrix {\cF^{-\Lambda} \cr
\cG^-_\Lambda\cr}\right)^\prime;    \quad
{\cal S}=\left( \matrix{A& B\cr C & D \cr}\right)\in G \subset Sp(2 n_v,\IR)
\end{equation}
 where:
 $
\cG^-_\Lambda = \bar \cN_{\Lambda\Sigma}\cF^{-\Sigma}$,
 $\cG^+_\Lambda =  ( \cG^-_\Lambda )^\star
$.
  $\cN_{\Lambda\Sigma}$ is the matrix appearing in the kinetic
  part of the vector Lagrangian:
  \begin{equation}
  \cL_{kin}= {\rm i}\bar \cN_{\Lambda\Sigma}\cF^{-\Lambda} \cF^{-\Sigma}
  + h. c.
  \end{equation}
Using a complex basis in the vector space of $Sp(2 n_v)$, we may
rewrite the
symplectic matrix in the following way:
\begin{equation}
{\cal S}
\to U={1 \over {\sqrt{2}}}\left(\matrix{f+ {\rm i}h & \bar f + {\rm i}
\bar h \cr  f- {\rm i}h & \bar f - {\rm i}
\bar h \cr}\right)
\end{equation}
    The requirement $ {\cal S} \in Sp(2 n_v, \IR)$ implies:
 \begin{equation}
\left\lbrace\matrix{{\rm i}(f^\dagger h - h^\dagger f) &=& \bfone \cr
(f^\dagger \bar h - h^\dagger \bar f) &=& 0\cr} \right.
\label{specdef}
\end{equation}
 $f$ and $h$  are coset representatives of $G$ embedded in
 $Sp(2 n_v, \IR)$ and can be constructed in terms of the $L$'s.
The kinetic matrix $\cN$ turns out to be:
\begin{equation}
\cN= hf^{-1}
\label{nhf-1}
\end{equation}
 and transforms projectively under duality rotations:
 \begin{equation}
\cN^\prime = (C+ D \cN) (A+B\cN)^{-1}
\end{equation}

  By using (\ref{specdef}), (\ref{nhf-1})  we find that
   $
   (f^t)^{-1} = {\rm i} (\cN - \bar \cN)\bar f
$.
 As a consequence, in the transformation law of gravitino (\ref{tragra})
 and gaugino (\ref{tragau})
 we have to substitute  the embedded inverse coset representative:
 $
(L_{\Lambda AB}, L_{\Lambda I}) \to  (f_{\Lambda AB}, f_{\Lambda I})
$.
In particular, the dressed graviphotons and matter vectors take the
symplectic invariant form:
\begin{eqnarray}
T^-_{AB}&=& f^\Lambda_{AB}(\cN - \bar \cN)_{\Lambda\Sigma}\cF^{-\Sigma}=
f^\Lambda_{AB} \cG^-_\Lambda - h_{\Lambda AB} \cF^{-\Lambda}  \\
  T^-_{I}&=& f^\Lambda_{I}(\cN - \bar \cN)_{\Lambda\Sigma}\cF^{-\Sigma}=
f^\Lambda_{I} \cG^-_\Lambda - h_{\Lambda I} \cF^{-\Lambda}
\end{eqnarray}
Obviously, when $N>4$ $L_{\Lambda I} = f _{\Lambda I}= T_I =0$.

 To construct the dressed charges one integrates
 $T_{AB} = T^+_{AB} + T^ -_{AB}  $ and  (for $N<4$)
 $T_I = T^+_I + T^ -_I $ on a large 2-sphere.
 For this purpose we note that
 \begin{eqnarray}
 T^+_{AB} & = & f^\Lambda_{AB} \cG_\Lambda^+  - h_{\Lambda
 AB}F^{+\Lambda} =0  \\
   T^+_I & = & f^\Lambda_{I} \cG_\Lambda^+  - h_{\Lambda
 I}F^{+\Lambda} =0
\end{eqnarray}
as a consequence of eq. (\ref{nhf-1}) and of the definition of $\cG^+$.
Therefore we have:
\begin{equation}
\pmatrix{Z_{AB}\cr Z_I} = \int \pmatrix{T_{AB}\cr T_I }
=  \int \pmatrix{T^ -_{AB}\cr  T^ -_I} =
\pmatrix{f^\Lambda_{AB}\cr f^\Lambda_I } e_\Lambda
- \pmatrix{h_{\Lambda AB}\cr h_{\Lambda I} } g^\Lambda
\label{zabi}
\end{equation}
where:
$
e_\Lambda = \int_{S^2} \cG_\Lambda $, $
g^\Lambda = \int_{S^2} F^\Lambda
$\\
 We see that the presence of dyons in $D=4$ is related to the
 symplectic embedding.
Also in this case one can obtain differential relations
and a sum rule among the charges.
The sum rule has the following form:
\begin{equation}
Z_{AB} \bar Z_{AB} + Z_I \bar Z_I = -{1\over 2} P^t \cM (\cN) P
 , \quad
P=\left(\matrix{g^\Lambda \cr e_ \Lambda \cr} \right)
 \end{equation}
where:
\begin{equation}
\cM(\cN) = \left( \matrix{ \bfone & 0 \cr - Re \cN &\bfone\cr}\right)
\left( \matrix{ Im \cN & 0 \cr 0 &Im \cN^{-1}\cr}\right)
\left( \matrix{ \bfone & - Re \cN \cr 0 & \bfone \cr}\right)
\label{m+}
\end{equation}

In order to obtain this result we just have to use the
 fundamental identities (\ref{specdef}) and    the definition of the
 kinetic matrix given in (\ref{nhf-1}).

Furthermore the Maurer--Cartan equations (\ref{dllp}) for the coset
representatives of $G/H$
imply analogous Maurer--Cartan equations for the embedded coset
representatives $(f,h)$:
\begin{equation}
\nabla^{(\IH)} (f,h) = (\bar f, \bar h)P^{(\IK)}. \label{mce}
\end{equation}
Using the definitions of the charges one finds the following
differential constraint:
\begin{equation}
\nabla^{(\IH)} Z_{(\IH)}\equiv\nabla (f^\Lambda_{(\IH)}e_\Lambda -
 h_{\Lambda (\IH)} g^\Lambda)= (\bar f^\Lambda_{(\IK)} e_\Lambda -
\bar h_{\Lambda (\IK)} g^\Lambda) \equiv \bar Z_{(\IK)} P^{(\IK)}
\end{equation}
Equation (\ref{mce}) can then be found in a way analogous to that shown
 before for the odd dimensional cases.
 \par
 For the other maximally extended even dimensional theories, there is
 an embedding problem analogous to the four dimensional case for the
 4--form in $D=8$ and for the 3--forms in $D=6$.  This is discussed
 in detail in \cite{adf}.\\
 For reasons of brevity, we do not analyze these cases here, since they
 do not present essential conceptual differences with respect
 to the $D=4$ case.
 Some modifications in fact occur in $D=6$,
 for in this case
 the Gaillard--Zumino duality group is an orthogonal group
 ($O(5,5)$) instead of $Sp(2n,\IR)$. Furthermore in $D=6$,  due to  the
 irreducible decomposition of a 3--form into self-dual and
 antiself-dual parts,
 there is no
 real
 distinction
 between electric and magnetic charges \cite{sagn}.


\newcommand{\Journal}[4]{{#1} {\bf #2}, #3 (#4)}
\newcommand{\NCA}{\em Nuovo Cimento}
\newcommand{\NIM}{\em Nucl. Instrum. Methods}
\newcommand{\NIMA}{{\em Nucl. Instrum. Methods} A}
\newcommand{\NPB}{{\em Nucl. Phys.} B}
\newcommand{\PLB}{{\em Phys. Lett.}  B}
\newcommand{\PRL}{\em Phys. Rev. Lett.}
\newcommand{\PRD}{{\em Phys. Rev.} D}
\newcommand{\ZPC}{{\em Z. Phys.} C}


\begin{thebibliography}{50}
\bibitem{sw}
N. Seiberg and E. Witten, Nucl. Phys. B426 (1994) 19;
Nucl. Phys. B431, 484 (1994)
\bibitem{ht}
C. Hull and P. K. Townsend hep--th/9595073; Nucl. Phys. B438 (1995) 109
\bibitem{wit}
E. Witten, Nucl. Phys. B443 (1995) 85
 \bibitem{sen}
 A. Sen,
Nucl. Phys. B440 (1995) 421; Phys. Lett. B303 (1993)  221;
Mod. phys. Lett.A10 (1995) 2081
\bibitem{ss}
J. Schwarz and A. Sen, Phys. Lett. B312 (1993) 105
 \bibitem{cdfv}
A. Ceresole, R. D'Auria, S. Ferrara and A. Van Proyen,
Nucl. Phys. B444 (1995)  92
 \bibitem{ht2}
C. Hull and P. K. Townsend, hep--th/9505073
\bibitem{w2}
E. Witten, hep--th/9507121
\bibitem{fk1}
S. Ferrara and R. Kallosh, Phys. Rev. D54 (1996)1514--1524;
Phys. Rev. D54 (1996) 1525--1534
 \bibitem{adf}
L. Andrianopoli, R. D'auria and S. Ferrara, hep-th/9608015
 \bibitem{sase}
A. Salam and E. Sezgin, ``Supergravities in diverse Dimensions''
Edited by A. Salam and E. Sezgin, North--Holland, World Scientific
1989, vol. 1
 \bibitem{adf2}
L. Andrianopoli, R. D'auria and S. Ferrara,
in preparation
\bibitem{cdf}
A. Ceresole, R. D'Auria and S. Ferrara, hep--th/9509160
\bibitem{cadafr}
L. Castellani, R. D'auria and P. Fre',
``Supergravity and Superstrings: a Geometric Perspective''
World Scientific 1991
 \bibitem{gz}
M. K. Gaillard and B.  Zumino, Nucl. Phys. B193  (1981)  221
 \bibitem{sagn}
S. Ferrara, R. Minasian and A. Sagnotti, hep-th/9604097

\end{thebibliography}
\end{document}